\documentclass{ws-procs975x65}
\def\be{\begin{eqnarray}}
\def\ee{\end{eqnarray}}

\def\tr{\textnormal{tr}}
\newcommand{\sbar}[1]{\ooalign{\hfil/\hfil\crcr$#1$}}

\begin{document}

\title{Dilatons  in Dense Baryonic Matter}

\author{Hyun Kyu  Lee$^*$ and Mannque Rho$^{\dagger}$ }

\address{$^*$Department of Physics, Hanyang University,
Seoul 133-791, Korea\\
E-mail: hyunkyu@hanyang.ac.kr \\
$^{\dagger}$ Institut de Physique Th\'eorique,
CEA Saclay, 91191 Gif-sur-Yvette c\'edex, France \\
Department of Physics, Hanyang University,
Seoul 133-791, Korea\\
 E-mail:mannque.rho@cea.fr}

\begin{abstract}
We discuss the role of dilaton, which is supposed  to be representing a special feature of scale symmetry of QCD,  trace anomaly, in dense baryonic matter. The idea that the scale symmetry breaking of QCD is responsible for the spontaneous breaking of chiral symmetry is presented  along the similar spirit of  Freund-Nambu model.  The incorporation of  dilaton field in the hidden local symmetric parity doublet model is briefly sketched with the  possible role of dilaton at high density baryonic matter, the emergence of linear sigma model in dilaton limit.
\end{abstract}


\bodymatter

\section{Introduction}

The recent discovery of Higgs-like particles at LHC activates  many interesting research not only on  phenomenology but also on the origin of Higs particle\cite{peskin,yamawaki}.  In the framework of Standard Model  the masses of leptons and quarks are explained by Higgs mechanism. But the  question on how masses of hadrons are generated has not been  satisfactorily  answered yet\cite{MRtalk}.    Since the light quark masses are too small to explain the hadron mass, the proton mass, for example, it is believed that the masses of hadrons are generated dynamically in the framework of quantum chromodynamics (QCD).

If we ignore  quark masses, a very good approximation for the hadrons consisting of light
quarks),  QCD has no scale at the classical level, but  a scale  emerges at the quantum level in the form of  trace anomaly, an explicit breaking of scale symmetry.   On the other hand, the spontaneous symmetry breaking of chiral symmetry($\chi$SB) gives rise to another important scale determined by the quark condensate, $\langle\bar{q}q\rangle$, which is supposed to generate most of the hadron masses. Thus it is natural to ask  whether the chiral symmetry breaking  and the scale symmetry breaking (SSB) are linked to each other and if so, how intricately.

The idea as to how $\chi$SB is tied to SSB has been discussed recently in connection to the discovery of Higgs-like  particles in LHC.  In the technicolor scheme, the {\it explicit}  $\chi$SB in heavy quark sector   triggers the spontaneous scale-symmetry breaking (SSB), of which a Nambu-Goldstone boson  is identified as a Higgs-like  particle.  It turns out that, in this walking technicolor scheme\cite{yamawaki}, the phenomenology of Goldston boson of  SSB can be differentiated from  Higgs in the standard model in more detailed analysis of LHC data.  And this development may shed new light on new mechanism of mass generation for hadrons.

In hadronic sector,  the idea that the scale symmetry and chiral symmetry are tied to each other has been explored in  the line of thought that the explicit breaking of scale symmetry due to the QCD anomaly triggers the spontaneous  $\chi$SB \cite{LR}.  Assuming that the QCD anomaly can be decomposed into two parts, ``hard" and ``soft,"  the  explicit symmetry breaking due to the soft part  can be treated as small enough so that there is an  approximate scale symmetry  at low energy scale. Then the dilaton emerges as a Nambu-Goldstone boson as argued in  the work of Freund and Nambu\cite{nambu}.

We can construct a  phenomenological Lagrangian  which implements the above idea by introducing dilaton field into the Lagrangian of pions and vector mesons constructed under the hidden local gauge symmetry principle.  One of the interesting consequences  is  the scaling of Lagrangian parameters: masses and coupling  constants are dialed by the change of vacuum expectation value of dilaton field, i.e., BR scaling\cite{BR}.
The possible forms of scaling laws have been explored in various contexts. One of them is
the observation of a half-skyrmion phase when skyrmions are put on the lattice which can be implemented into a new scaling law giving  a stiffer equation of state\cite{LPR, dongetal}.  Another interesting feature of dilaton field is  the dilaton limit, that leads to the condition that  a particular combination of pions and dilaton lead to the linear sigma model  giving rise to a  nontrivial scaling behavior\cite{bira, paeng}.

In section 2, the work of Freund and Nambu is briefly sketched to develop a model in which pions  and dilaton are introduced to take care of  spontaneous breaking of both chiral symmetry and approximate scale symmetry. We introduce the phenomenological Lagrangian, denoted dHLS,  which implements the above idea by introducing dilaton field into the Lagrangian constructed under the  hidden local gauge symmetry principle\cite{HY:PR} .  In section 3, we elaborate on  how a dilaton field  can be applied  to the entangled transformation of  pions in  nonlinear realization of $\chi$SB at lower scale such that the chiral symmetry is linearly  realized  in the form of  linear sigma model\cite{bira}. The summary is given in section 4.

\section{Dilaton and Chiral Symmetry Breaking}

The first example which demonstrates the possibility of spontaneous symmetry breaking of the scale symmetry and its physical application of the dilaton field is the Freund-Nambu model\cite{nambu, Zumino}. It  has two real scalar fields, $\psi$ and $\phi$, with the potential
 \be
V(\psi, \phi) = V_a + V_b,\label{fn}\ee
where \be V_a &=& \frac{1}{2} f^2 \psi^2 \phi^2, ~~~
V_b = \frac{\tau}{4}[\frac{\phi^2}{g^2} - \frac{1}{2} \phi^4
  - \frac{1}{2g^4}]= \frac{\tau}{8g^4}(g^2\phi^2 -1)^2 \label{fnsb}.
  \ee
$V_a$ is scale invariant but  $V_b$ includes scale symmetry breaking terms $\frac{\phi^2}{g^2}$ and $- \frac{1}{2g^4}$.
One can introduce the new field, $\chi$, defined by
 \be
\chi = (g^2\phi^2 -1)/(2g),\label{chi}
 \ee
of which the scale transformation
\be
\delta \chi = \epsilon(x \cdot \partial +2) \chi +
\frac{\epsilon}{g},\label{gb}
 \ee
manifests one of the characteristics of Nambu-Goldstone bosons. In terms of the condensate $\phi_0$ given by the minimum of the potential, $\phi_0=1/g$,
the masses take the form
 \be
 m^2_{\psi} =f^2\phi_0^2, ~~~~ m^2_{\chi} = \tau\phi_0^2.
 \ee
It should be noted that the mass of $\psi$ comes from the scale invariant term $V_a$,  so is independent of $\tau$. Thus it can
have an arbitrary value.  However $m_{\chi}$ depends on $\tau$, going to zero linearly as
$\tau\rightarrow 0$.
This is known to be a characteristic of an approximate
spontaneously-broken scale symmetry\cite{Zumino}.

Let us consider the Lagrangian of linear sigma model with $\Phi = \sigma + i \tau \cdot \vec{\phi}$ with a potential of the following form
\be
V(\phi) =\lambda (\frac{1}{2} Tr \Phi \Phi^{\dagger}  -v^2)^2
= \lambda(\sigma^2 + \Sigma_i \phi^2_i -v^2)^2. \label{vphi}
\ee
The scale invariance is broken due to nonvanishing $v^2$.  But it  is also this term which breaks chiral symmetry spontaneously.  When we assign  the  vacuum expectation  only to $\sigma$,    $\phi_i, i = 1,2,3$ become Nambu-Goldstone bosons  related to the spontaneous chiral symmetry breaking and we call them pions, $\pi$.   For $\lambda=0$, there is no explicit breaking of scale symmetry regardless of the value of $v^2$.
We introduce a scalar  field, $\chi$,  and reformulate the above potential in the following form,
\be
V(\phi, \chi) = \lambda (\frac{1}{2} Tr \Phi \Phi^{\dagger} -\chi^2)^2 + V(\chi) \label{vphichi}
\ee
The first term is analogous to  $V_a$ in eq.(\ref{fnsb}), which is scale invariant. The second term is to break scale symmetry.
In general, $V_{\chi}$ gives nonvavinshing vacuum expectation value for $\chi$, which can be of the following forms
\be
V_{\chi}= \frac{\tau}{8}(\chi^2 - v^2)^2,  ~~~
or ~~~
V_{\chi}= \tau \chi^4 \ln \frac{\chi}{e^{1/4}v}\,.\label{vphichiln}
\ee
 One can see that the scale symmetry breaking in eq.(\ref{vphi}) is  transferred to the second term of eq.(\ref{vphichi}), which plays a similar role as  eq.(\ref{fnsb}).
 If the scale symmetry breaking term  with $\tau$ is small enough such that there is an {\it approximate} scale symmetry,  then it can be considered as spontaneously broken and associate a Nambu-Goldstone boson of the form\cite{Zumino, sz1},
\be
\eta = (\chi^2 - v^2)/2v, ~~~  or ~~~
   \chi   &=& v e^{\eta/v}.
\ee

The origin of scale symmetry breaking in hadron physics can be traced back to the QCD trace anomaly and a dilaton field can be associated with the process.
Then eq.(\ref{vphichi}) reduces to eq.(\ref{vphi}).
It is now clear in this formulation that the spontaneous chiral symmetry breaking is not possible without scale symmetry breaking. Or put differently, {\it scale symmetry breaking is responsible for the chiral symmetry breaking}.  This indicates one of the possible ways that  the trace  anomaly of QCD  can  be linked to chiral symmetry breaking.

The chiral symmetry, which is broken spontaneously,  can be realized non-linearly by introducing $U(\pi) =  e^{i~\vec{\tau}\cdot \vec{\pi}/ F_{\pi}}$ instead of  four-component scalar fields, $\Phi$. The pions $\pi_i, ~~ i = 1,2,3$ transform nonlinearly under the broken sector of chiral transformation, with the pions derivatively coupled.  The lowest derivative term of pion dynamics (namely the current algebra term) can be written
\be
{\cal L}= \frac{F_{\pi}^2}{4} \mbox{tr}\left[ \partial_\mu U \partial^\mu U^\dagger \right].
\ee
This term breaks scale symmetry explicitly  because of the explicit scale $F_{\pi}$. This scale can be traced back to  the vacuum expectation value of the scalar field $\chi$ as in  eq.(\ref{vphichi}) with $F_{\pi} = v$ and the scale symmetry breaking can be traded in the  potential  $V(\chi)$
\be
{\cal L}(U, \chi)= \frac{\chi^2}{4} \mbox{tr}\left[ \partial_\mu U \partial^\mu U^\dagger \right] + V(\chi) + \frac{1}{2} \partial_\mu \chi \partial^\mu \chi.
\ee

Defining a new field
\be
\Sigma  \equiv \chi U(\pi) , \label{linear}
\ee
we can see that
\be
{\cal L}(U, \chi)  \rightarrow {\cal L}(\Sigma) =\frac{1}{4} \mbox{tr} \left[ \partial_\mu \Sigma \partial^\mu \Sigma^\dagger \right] + V(\Sigma\Sigma^{\dagger}),\label{linear2}
\ee
where we have used $\chi = \frac{1}{2}Tr\Sigma \Sigma^{\dagger}$.  If we can identify  $\Sigma$ as
\be
\Sigma = \sigma + i\vec{\pi}\cdot \vec{\tau},
\ee
then it is nothing but  the linear sigma model.  However it is not clear how
the linear representation of nonlinear field can be  achieved by adding a dilaton, which is  chiral singlet.  Apart from the theoretical justification, we  may expect this  can be achieved physically in a certain limit, for example in  baryonic matter with higher density.  This is the issue that concerns the ``dilaton limit" discussed in\cite{bira,paeng}.

\section{Dilton Limit of dHLS}

It has been observed that the dilaton limit leads to a  very interesting consequence when  nucleons are included in the chiral Lagrangian.  The vector coupling constant is found to  have a limiting value $g \rightarrow 1$ in the dilaton limit.  This idea has been extended\cite{paengKRS} to the effective theory where dilaton is incorporated into the  hidden local symmetry theory, where vector mesons are present as explicit degrees of freedom.  In this section the results are briefly sketched.

The dilaton field which is responsible for the spontaneous symmetry breaking of chiral symmetry, thus generating pions, can be naturally included in the effective Lagrangian at low baryon density.  The redundant symmetry, i.e., hidden local symmetry(HLS), appearing in the formalism of nonlinear realization of spontaneous chiral symmetry breaking  has been used to incorporate the vector mesons systematically.  However  the Lagrangian of HLS is, by construction,  noninvariant under scale symmetry.   The pattern  of symmetry breaking is thus not consistent with the scale symmetry breaking due to QCD trace anomaly, which is believed to be the only reason why the effective theory of QCD should have scale symmetry breaking term modulo the light quark mass.     The dilated HLS is constructed such that all terms in the Lagrangian are scale invariant except for the potential of dilaton, that reproduces the relevant scale symmetry breaking for QCD trace  anomaly.   One of the simplest ways to construct the dilated HLS is
to follow the standard trick of inserting the dilaton field  $\chi$ as  ``conformal compensator"  into the Lagrangian to  obtain scale symmetric Lagrangian, given by
\begin{eqnarray}
{\cal L} &=& {\cal L}_N + {\cal L}_M + {\cal L}_\chi\,,
\label{dlfplag} \\
\mathcal{L}_{N}
&=& \bar{Q}i\gamma^{\mu}D_{\mu}Q - g_{1}F_{\pi}\frac{\chi}{F_{\chi}}\bar{Q}Q
{}+ g_{2}F_{\pi}\frac{\chi}{F_{\chi}}\bar{Q}\rho_{3}Q
{}- im_{0}\bar{Q}\rho_{2}\gamma_{5}Q
\nonumber\\
&&+ g_{V\rho} \bar{Q}\gamma^{\mu}\hat{\alpha}_{\parallel \mu}Q
{}+ g_{V0} \bar{Q}\gamma^{\mu}\mbox{tr}\left[\hat{\alpha}_{\parallel \mu} \right]Q
{}+ g_{A}\bar{Q} \rho_{3} \gamma^{\mu}\hat{\alpha}_{\perp \mu}
\gamma_{5} Q\,,
\label{NchiLargrangian} \\
{\mathcal L}_M
& = & \frac{F_{\pi}^2}{F_{\chi}^2} \chi^2\mbox{tr}\left[ \hat{\alpha}_{\perp\mu}
  \hat{\alpha}_{\perp}^{\mu} \right]
{}+ \frac{F_{\sigma\rho}^2}{F_{\chi}^2} \chi^2\mbox{tr}\left[ \hat{\alpha}_{\parallel\mu}
  \hat{\alpha}_{\parallel}^{\mu} \right]
{}+ \frac{F_{\sigma\omega}^2 - F_{\sigma\rho}^2}{2F_{\chi}^2} \chi^2\mbox{tr}\left[ \hat{\alpha}_{\parallel\mu} \right]
  \mbox{tr}\left[ \hat{\alpha}_{\parallel}^{\mu} \right] \nonumber \\
&& {}- \frac{1}{2}\mbox{tr}\left[ \rho_{\mu\nu}\rho^{\mu\nu} \right]
{}- \frac{1}{2}\mbox{tr}\left[ \omega_{\mu\nu}\omega^{\mu\nu} \right]\,,
 \\
{\mathcal L}_\chi
&=& \frac{1}{2}\partial_\mu\chi\cdot\partial^\mu\chi {}-V(\chi)
\end{eqnarray} where $V(\chi)$ is the  dilaton potential that breaks scale symmetry spontaneously and
$F_{\chi}$ is the vacuum expectation value of $\chi$ at zero temperature and density.
The detailed expressions needed in this section can be found in ref.  \cite{paengKRS} .
As in the previous section, we do the field re-parameterizations $\Sigma_s=U\chi\frac{F_\pi}{F_\chi}=s+i\vec{\tau}\cdot \vec{\pi}$. And using the nonlinearly transforming nucleon field of parity eigenstates,
 one finds a complicated expression for (\ref{dlfplag}) composed of a part that is regular, ${\cal L}_{\rm reg}$, and a part that is singular, ${\cal L}_{\rm sing}$, as $\mbox{tr}(\Sigma_s\Sigma_s^\dagger)= 2\left( s^2 + \pi^{a\,2}\right)
\rightarrow 0$, where $a$ is iso-spin index.
The singular part that arises solely from the scale invariant part of the
original Lagrangian (\ref{dlfplag}) has the form
\begin{eqnarray}
\mathcal{L}_{\rm sing} =
\left( g_{V\rho} -g_A \right) {\cal A} \left( 1/\tr \left[ \Sigma_s \Sigma_s^{\dagger} \right]\right) + \left( \alpha -1\right) {\cal B} \left( 1/\tr \left[ \Sigma_s \Sigma_s^{\dagger} \right]\right)\,,
\label{sing}
\end{eqnarray}
One can see that the condition  $\alpha \equiv \frac{F_\pi^2}{F_\chi^2} =1$ has been  already used in getting Eq.(\ref{linear2}).
That ${\mathcal L}_{\rm sing}$ be absent also leads to the condition that
\be
g_{V\rho}-g_A\rightarrow 0\,.
\ee

Using large $N_c$ sum-rule arguments~\cite{bira}
and the RGE \cite{paeng}, we infer
\be
g_A-1\rightarrow 0\,.
\ee
It follows then that
\be
g_{\rho NN}=g_\rho(g_{V\rho}-1)\rightarrow 0.
\ee
This set of limits defines what is referred to as  ``dilaton limit."
We thus find that in the dilaton limit, the $\rho$ meson decouples from the nucleon. In contrast, the limiting $\mbox{tr}(\Sigma_s\Sigma_s^\dagger)\rightarrow 0$ {\em does not} give any constraint on $(g_{V\omega}-1)$. The $\omega$-nucleon coupling remains non-vanishing in the Lagrangian of fluctuations $\tilde{s}$ and $\tilde{\pi}$
around their expectation values, which in terms of the mass-diagonalizing field  ${\cal N}$,  takes the form ,
\begin{eqnarray}
{\mathcal L}_N
=
&\bar{\cal N}i\sbar{\partial}{\cal N} - \bar{\cal N}\hat{M}{\cal N}
{}- g_1\bar{\mathcal N}\left(
\hat{G}\tilde{s} + \rho_3\gamma_5 i\vec{\tau}\cdot\vec{\tilde{\pi}}
\right) {\mathcal N}
\nonumber\\
&
{}+ g_2\bar{\mathcal N}\left(
\rho_3 \tilde{s} + \hat{G}\gamma_5 i\vec{\tau}\cdot\vec{\tilde{\pi}}
\right) {\mathcal N}
+ \left(1-g_{V\omega} \right) g_\omega{\cal N}  \frac{\sbar{\omega}}{2}  \mathcal{N} .
\end{eqnarray}
 This is just the nucleon part of the linear sigma model in which the $\omega$ is minimally coupled to the nucleon.

\section{Summary}

The origin of scale symmetry breaking in hadron physics can be traced back to the QCD trace anomaly, and it can be implemented into an effective theory by introducing  a dilaton field. The idea that the scale symmetry and chiral symmetry are tied to each other has been explored in  a line of thought that the explicit breaking of scale symmetry due to the QCD anomaly triggers the spontaneous  $\chi$SB \cite{LR}. We demonstrate that using a model similar to that of Freund-Nambu the spontaneous chiral symmetry breaking is locked to the  scale symmetry breaking through dilaton field.  This provides one of the possible ways to relate the trace  anomaly of QCD  to chiral symmetry breaking.

It is possible to construct a phenomenological Lagrangian\cite{paengKRS}  that encapsulates the above idea by introducing dilaton field into the Lagrangian of pions and vector mesons constructed under the hidden local gauge symmetry principle with parity doubled baryons.  One of the interesting consequences  is  the scaling of parameters of the Lagrangian: masses and coupling  constants are dialed by the change of vacuum expectation value of  dilaton field as in BR scaling \cite{BR,dongetal}.
Another interesting feature of dilaton field is  the dilaton limit that gives the condition that a particular combination of pions and dilaton lead to the linear sigma model,  giving rise to a  nontrivial scaling behavior \cite{bira, paeng}.
 Most of the conventional effective actions constructed for low-density phenomena without dilaton available in the literature do not reveal  the above feature.  To realize the linear realization feature,  the coupling constants should satisfy the particular condition.  A highly non-trivial support for the dilaton limit properties comes from a renormalization group analysis of hidden local symmetry implemented with baryons\cite{paeng}.  An intriguing consequence of the analysis is that the $\rho$ meson nucleon coupling vanishes at critical density corresponding to the dilaton limit {\it before} reaching chiral restoration.

\section*{Acknowledgments}
 HKL would like to thank Koichi Yamawaki for kind invitation to SCGT 12, Nagoya  and for hospitalities during the conference. He also would like to  thank Mannque Rho,  Masayasu Harada, Tom Kuo,  Won-Gi Paeng,  Byung-Yoon Park and Chihiro Sasaki for helpful discussions.  This work is supported by WCU (World Class University) program: Hadronic Matter under Extreme Conditions through the National Research Foundation of Korea funded by the Ministry of Education, Science and Technology (R33-2008-000-10087-0).


\end{document}